\documentclass{article}

\usepackage{PRIMEarxiv}

\usepackage[utf8]{inputenc} 
\usepackage[T1]{fontenc}    
\usepackage{hyperref}       
\usepackage{url}            
\usepackage{booktabs}       
\usepackage{amsfonts}       
\usepackage{nicefrac}       
\usepackage{microtype}      
\usepackage{lipsum}
\usepackage{fancyhdr}       
\usepackage{graphicx}       
\graphicspath{{media/}}     
\usepackage{array,multirow}
\usepackage{graphicx}
\usepackage{dcolumn}
\usepackage{bm}
\usepackage[version=4]{mhchem}
\usepackage{comment}
\newcommand{\mi}{{\rm i}}

\newcommand{\rr}{{\bm r}}
\newcommand{\ee}{{\bm e}}
\newcommand{\kk}{{\bm k}}
\newcommand{\GG}{{\bf G}}

\newcommand{\GGp}{{\bf G}^{(1)}_{\rm pl}}

\newcommand{\IM}{{\rm{Im}}\,}

\newcommand{\pH}{\textrm{p}\ce{H}}
\newcommand{\pK}{\textrm{p}K}

\usepackage[utf8]{inputenc}
\usepackage[T1]{fontenc}
\usepackage{mathptmx}
\usepackage{etoolbox}
\usepackage{xcolor}
\usepackage{}
\usepackage{tabularx}
\title{pH-sensitive spontaneous decay of functionalised carbon dots in solutions
}

\author{
  Denise Dilshener \\
  Department of Physics and Technology  \\
  University of Bergen \\
  Bergen, Norway\\
   \And
  Drew F. Parsons\\
  Department of Chemical and Geological Sciences \\
  University of Cagliari \\
  Cagliari, Italy
 \AND
 Johannes Fiedler \\
  Department of Physics and Technology  \\
  University of Bergen \\
  Bergen, Norway\\
  \texttt{johannes.fiedler@uib.no} \\
}

\begin{document}
\maketitle

\begin{abstract}
Carbon quantum dots have become attractive in various applications, such as drug delivery, biological sensing, photocatalysis, and solar cells. Among these, pH sensing via luminescence lifetime measurements of surface-functionalised carbon dots is one application currently investigated for their long lifetime and autonomous operation. In this manuscript, we explore the theoretical connection between excitation lifetimes and the pH value of the surrounding liquid via the protonation and deprotonation of functional groups. Example calculations applied to m-phenylenediamine, phloroglucinol and tethered disperse blue 1 are shown by applying a separation approach treating the electronic wavefunction of functional groups separately from the internal electronic structure of the (large) carbon dot. The bulk of the carbon dot is treated as an environment characterised by its optical spectrum that shifts the transition rates of the functional group. A simple relationship between \pH, p$K_a$ and mixed fluorescence lifetime is derived from transition rates of the protonated and deprotonated states. \pH\ sensitivity improves when the difference in transition rates is greatest between protonated and deprotonated species, with the greatest sensitivity found where the p$K_a$ is close to the pH region of interest. The introduced model can directly be extended to consider multicomponent liquids and multiple protonation states.
\end{abstract}


\section{Introduction}
\pH\ is an essential factor in a variety of scientific and industrial fields. A slight variation in pH can significantly influence all processes and systems governed by pH levels. In aquaculture, specifically fish farming, monitoring water quality is vital. When transporting live fish in wellboats from one location to another, it is crucial to ensure they are not exposed to potentially harmful environmental conditions like algal blooms or oil spills.\cite{tang2009modeling} This necessitates careful observation of water conditions to prevent the wellboat from circulating lower-quality, contaminated water.\cite{thomas2017use}

Ocean acidification, a process resulting in lowered pH levels, significantly alters marine environments' carbonate chemistry. \cite{das2015ocean} This profoundly affects the marine microbiome, which plays a central role in ocean ecosystems. Therefore, it's essential to employ advanced technologies to study the effects of pH shifts on the microbiome to understand and mitigate against these changes.

Understanding and measuring pH across various applications highlights the need for advanced pH sensor technologies. Traditional pH electrodes have limitations; they need frequent calibration due to changes in their parameters over time, and they also require regular replenishment of electrolytes to compensate for consumption. Additionally, these electrodes often perform poorly in environments with high salinity, which can be a significant drawback in applications like marine research or industries dealing with brackish or seawater. Therefore, there is a demand for more advanced and reliable pH measurement solutions that can overcome these challenges.\cite{szapoczka2023fluorescence,wiora2018over}

Innovative pH measurement methods like ion-sensitive field-effect transistors (ISFETs), spectrophotometric systems, and optodes have been introduced to address the disadvantages of conventional pH electrodes. Yet, these technologies face challenges, including sensitivity to light and pressure, higher costs and power demands, and the need for specialised materials to ensure long-term stability. Moreover, accurately controlling temperature and ionic strength during seawater analysis, which significantly affects sensor performance, remains an unresolved issue with these advanced methods.\cite{staudinger2019fast}

Optical chemical sensors have gained attention due to their low cost, low power consumption and good long-term stability. When the pH in the surrounding liquid changes, the optical characteristics of the pH-sensing material change, which can be measured by detecting the fluorescence intensity. The challenge here is photobleaching from sunlight or probe light, leaching of the indicator from the immobilising medium and noise contributions from background luminescence, and probe light intensity variations.\cite{totland20209}

For these reasons, fluorescence lifetime is an interesting factor to observe using optical pH sensors, as it is an intrinsic characteristic of the material and can be a basis for more reliable sensors with long-term calibration stability.  It is not affected by the challenges addressed with fluorescence intensity.\cite{totland2019broad}

Previous experimental studies on this question have found acridine in an amine-modified silica material exhibits a remarkable fluorescence lifetime shift with pH shift with a linear decrease by a factor of $1/3$ over a pH range from 2 to 12 (20 ns between high and low pH).\cite{totland20209} However, chloride ions quench acridine fluorescence, so hence this system is unsuitable for marine applications.

Carbon dots (CDs) are of great interest in searching for a more suitable material. Carbon dots are highly luminescent, small-sized materials with low toxicity and good biocompatibility, sourced affordably for use in biomedical, catalytic, electronic, and security applications. Their fluorescence 
and easy dispersion in water makes them ideal for optical sensors due to their rapid and sensitive response. Their small scale with a high functionalised surface area also makes them highly responsive to environmental changes such as pH that affect their optical properties and fluorescence behaviour.\cite{liu2020carbon}

Carbon dots (CDs) consist of a carbon core with attached functional groups or dye molecules, and their fluorescence is thought to arise from quantum confinements or surface functionalization. The latter effect causes pH sensitivity, as considered in this manuscript. Functionalising CD surfaces with different dye molecules is a key method for creating sensors.\cite{liu2020carbon, bagheri2019carbon} Improved structural models will help identify optimal conditions for fluorescent CD applications. 

Previous studies synthesised CDs from various precursors and analysed fluorescence intensity (FI) and lifetime (FL) in pH 5-9 solutions.
FI varied with the precursor, while FL altered by about $0.5 \pm 0.2\, \mathrm{ ns}$ across pH levels depending on the precursor used.\cite{szapoczka2023fluorescence}

This work presents a theory to understand the pH dependence of the fluorescence lifetime of the different dye molecules functionalising a carbon dot's surface. 
The \pH\ sensitivity of the carbon dot will be active at the interface of the particle with the aqueous medium, hence we focus on describing responses of an attached dye molecule. For simplicity, we assume the response of the molecule to be unaffected by its chemical bond to the surface.
We derive a relationship based on quantum chemical FLs of protonated and deprotonated states of surface functional groups. The theory is 
demonstrated by applying to
the same dye molecules used in the work of W. Szapoczka \textit{et al.}: m-phenylenediamine (mPD), phloroglucinol and disperse-blue1 dye.\cite{szapoczka2023fluorescence,bagheri2019carbon}

\section{The impact of pH on spontaneous decays}
The quantum-mechanical structure of carbon dots yields photoluminescent properties.~\cite{Zhang2021} Regarding quantum electrodynamics, this effect is associated with real photons located in the visible spectrum. In contrast, pH is a measure of the \ce{H+} ion concentration, meaning pH affects electrostatic forces. Hence, the optical impact of pH on the decay of a given molecule is negligible.~\cite{C9CP03165K} Thus, a further intervening mechanism is required to use carbon dots for pH sensing: the protonation and deprotonation of functional groups, leading to changed electronic structures and providing a distinguishable signal. Consequently, these functional groups need to be located on the surface of the carbon dot to be chemically active for protonation and deprotonation. Following these thoughts, we consider the dye centre of the precursor dye molecules used during the synthesis of the carbon dot to functionalise its surface, thus carrying the primary responsibility for the pH sensitivity. The impact of the hosting carbon dot and the solvent medium on the decay is considered in line with the Purcell effect~\cite{PhysRev.69.37,Scheel2008,10.1063/5.0106503}. In the following, we will derive a model mapping the pH value on averaged decay rates for a mixture of protonated and deprotonated functionalised carbon dots. We then calculate the impact of the neighbouring environment of the carbon dot and solvent on the molecule fluorescence lifetime (and excitation energy) via coupling the functional group's properties to the dressed quantum field of the environment. Finally, we present the pH-sensitive excitation lifetimes based on dye properties calculated via DFT and TDDFT.
\subsection{Protonation and deprotonation of dyes and its impact on transition rates}\label{sec:mixing}

The pH sensitivity in the carbon dot's fluorescence response arises from the pH-dependent partitioning between protonated and deprotonated forms of dye molecules embedded in the carbon dot. 
We represent a protonated dye species by symbol \ce{HD} and deprotonated by symbol \ce{D}. Note that \ce{D} may be neutral or charged, depending on the system (charge $-1$ for the acidic hydroxyl groups of phloroglucinol; charge 0 for the basic amine groups of m-PD). The deprotonated fraction of the total number of dye molecules active on the surface of the carbon dot is $f_{\rm D}({\rm pH})$. The fraction of protonated dye molecules is then $f_{\ce{HD}}({\rm{pH}}) = 1 - f_{\ce{D}}$.  In our model, we assume the fluorescence transition rate of the carbon dot is a linear combination of the transition rates of the protonated and deprotonated forms of the dye molecules bound to the carbon dot such that
\begin{equation}
    \Gamma(\pH) = f_{\ce{D}}(\pH)\, \Gamma_{\ce{D}} + [1 -  f_{\ce{D}}(\pH)]\,  \Gamma_{\ce{HD}}\,.\label{eq:GammaPH}
\end{equation}
The chemical equilibrium between the two forms of the dye molecule (not calling into consideration the charge of \ce{D}, which, as noted above, may be 0 or $-1$ depending on whether the dye is acidic or basic) can be written as
\begin{equation}
\ce{HD ->[K_a] H+ + D}    \,,
\end{equation}
with acid equilibrium constant $K_a$ typically written in log form $\mathrm{p}K_{a} = -\log_{10} K_{a}$. The equilibrium is commonly presented in reference to pH via the Henderson–Hasselbach equation\cite{WencelAbelMcDonagh2014}
\begin{equation}
    \pH = \mathrm{p}K_{a} + \log_{10} \frac{\ce{[D]}}{\ce{[HD]}}\,.
\end{equation}
Identifying that $\ce{[D]}/\ce{[HD]} = f_{\rm D}/(1-f_{\rm D})$, we may write the dye fractions as
\begin{eqnarray}
    f_{\rm D} &=& \frac{1}{1 + 10^{\mathrm{p}K_{a}-\pH}}\,, \\
    f_{\rm HD} &=& \frac{1}{1 + 10^{\pH-\mathrm{p}K_{a}}}\,,
\end{eqnarray}
giving the pH-dependent transition rate of the carbon dot
\begin{equation}
    \Gamma(\pH) = \frac{\Gamma_{\rm HD} + \Gamma_{\rm D}\, 10^{\pH-\mathrm{p}K_{a}}}{1 +10^{\pH-\mathrm{p}K_{a} }}\,,\label{eq:ratemixing}
\end{equation}
which can be linearised  with a Taylor series expansion if $\mathrm{p}K_{a}$ lies close to the \pH\ ($\pH \approx \mathrm{p}K_a$) region of interest, as
\begin{equation}
    \tau(\pH) = \frac{2}{\Gamma_{\rm HD} + \Gamma_{\rm D}} +\frac{\ln 10\left(\Gamma_{\rm HD}-\Gamma_{\rm D}\right)}{\left(\Gamma_{\rm HD} + \Gamma_{\rm D}\right)^2} \left(\pH -\mathrm{p}K_a\right) \,. \label{eq:lin}
\end{equation}
Thus, we see that \pH\ sensitivity $\mathrm{d}\tau/\mathrm{d}\pH$ would be increased by maximising the difference in the transition rates of the two species ($\Gamma_{\rm HD}-\Gamma_{\rm D}$), while maintaining transition rates (in the sense of $\Gamma_{\rm HD}+\Gamma_{\rm D}$) as small as possible. In other words, greater sensitivity will be achieved the higher the fluorescence lifetime, subject to having a sufficient difference between the individual fluorescence lifetimes of each photoactive species.
\subsection{Medium-assisted transition rates}\label{sec:internal}
To determine the impact of the surrounding carbon and aqueous medium on the transition rate, the internal state dynamics of a particle in the presence of dielectric bodies have to be obtained. Thus, we separate the system~\cite{Scheel2008} into the the molecular system $\hat{H}_{\rm M}$; the electromagnetic fields $\hat{H}_{\rm F}$; and the molecule-field coupling $\hat{H}_{\rm MF}$,
\begin{equation}
    \hat{H} = \hat{H}_{\rm M} + \hat{H}_{\rm F} +\hat{H}_{\rm MF}\,. \label{eq:Ham}
\end{equation}
This equation demonstrates the limit of validity of the model,
namely that it is solely applicable to separable systems, meaning, in our case, that the colour centre, responsible for pH sensitivity, is separated from the host material. The quantification of this separability can occur by obtaining the overlap of the corresponding wave functions.~\cite{D0CP02863K,D2CP03349F}
Molecular states $\left|n\right\rangle$ are described by an infinite set of discrete wave functions leading to the diagonalised Hamiltonian
\begin{equation}
    \hat{H}_{\rm M} = \sum_n E_n \left|n\right\rangle\left\langle n\right| =\sum_n E_n \hat{A}_{nn}\, ,\label{eq:HamAtom}
\end{equation}
with the flip operator $\hat{A}_{mn} = \left|n\right\rangle\left\langle m\right|$. The molecule is coupled in the dipole approximation~\cite{Buhmann12a}
\begin{equation}
    \hat{H}_{\rm MF} = -\hat{\bm{d}}\cdot \hat{\bm{E}}({\bm{r}}_{\rm M}) =
    -\sum_{n,m} \hat{A}_{mn} {\bm d}_{mn} \cdot \hat{\bm{E}} ({\bm r}_{\rm M})\, ,
\end{equation}
with the molecule's position ${\bm{r}}_{\rm M}$. The electromagnetic field is described by the field Hamiltonian~\cite{Buhmann12a}
 \begin{equation}
     \hat{H}_{\rm F} = \sum_{\lambda=\rm{e,m}}\int\mathrm d^3 r \int\limits_0^\infty \mathrm d \omega \, \hat{\bm{f}}^\dagger_\lambda ({\bm r},\omega)\cdot \hat{\bm{f}}_\lambda({\bm r},\omega) \, ,
 \end{equation}
with the field's ladder operators $\hat{\bm{f}}_\lambda$ and $\hat{\bm{f}}^\dagger_\lambda$.

Heisenberg's equations of motion describe the dynamics of the molecular states~\cite{Buhmann12b}
\begin{eqnarray}
    \frac{d}{dt}{\hat{A}}_{mn} = \frac{1}{\mi\hbar}\left[ \hat{A}_{mn},\hat{H}\right] = \mi \omega_{mn}\hat{A}_{mn} +\frac{\mi}{\hbar}\sum_k \left(\hat{A}_{mk} {\bm{d}}_{nk} - \hat{A}_{kn}{\bm{d}}_{km}\right) \cdot \hat{\bm{E}}(\rr_{\rm A}) \, .\label{eq:motion}
\end{eqnarray}
Instead of explicitly solving this system of equations~(\ref{eq:motion}), the coupling matrix is diagonalised~\cite{Buhmann12b} and split into its imaginary and real part to determine the molecular frequency shifts for the $n$th excited state~\cite{Ribeiro2015}
\begin{equation}
    \delta\omega_{kn} = -\frac{\mu_0}{\hbar \pi} \mathcal{P} \int\limits_0^\infty\mathrm d \omega \frac{\omega^2 {\bm{d}}_{nk} \cdot {\rm{Im}} \, {\bf{G}}(\rr_{\rm A},\rr_{\rm A},\omega)\cdot {\bm{d}}_{kn}}{\omega+\omega_{kn}} \, ,\label{eq:deltaomega}
\end{equation}
and the corresponding transition rate~\cite{10.1063/5.0106503}
\begin{equation}
    \Gamma_n = \frac{2\mu_0}{\hbar}\sum_{k<n}  \omega_{nk}^2 {\bm{d}}_{nk} \cdot {\rm{Im}} \, {\bf{G}}(\rr_{\rm{A}},\rr_{\rm{A}},\omega_{nk})\cdot {\bm{d}}_{kn} \, . \label{eq:Gamma}
\end{equation}
The optical mode density is determined by the Green function as the general solution of the vector Helmholtz equation~\cite{Buhmann12a}
\begin{equation}
\left[ \nabla\times\frac{1}{\mu({\rr},\omega)}\nabla\times -\frac{\omega^2}{c^2}\varepsilon(\rr,\omega)\right] {\bf{G}}(\rr,\rr',\omega) = \boldsymbol{\delta}(\rr-\rr') \, ,   \label{eq:Green}
\end{equation}
determining the impact of the surrounding environment. Inserting the free-space Green function, $\rm{Im}\,\GG(\rr,\rr,\omega) = \omega/(6\pi c) {\bf{I}}$,~\cite{Scheel2008} one obtains the well-known Lamb-shift~\cite{Lamb1947}
\begin{equation}
    \Delta E = \frac{\mu_0}{6\pi^2 c}\sum_k \omega_k^3\left|{\bm{d}}_{0k}\right|^2 \ln\left(\frac{m_ec^2}{\hbar\omega_k}\right)\,,
\end{equation}
and Einstein coefficient~\cite{10.1119/1.12937} or Fermi's Golden rule~\cite{Dung2003,D2CP03349F}
\begin{equation}
    \Gamma_{nm} = \frac{\omega_{nm}^3 \left|{\bm{d}}_{nm}\right|^2}{3\hbar\pi\varepsilon_0c^3}\,.\label{eq:Einstein}
\end{equation}
As the optical mode-density $\rm{Im}\,\GG(\rr,\rr,\omega)$ can take positive and negative values, both blue and red detuned processes can be described.~\cite{PhysRevA.60.1590}
\subsection{Optimal parameters}
From the dependence of the transition rates on pH, Eq.~\eqref{eq:ratemixing}, an optimal parameter range for high pH sensitivity can be derived. Introducing the ratio between the deprotonated and protonated rates $\gamma = \Gamma_{\rm D}/\Gamma_{\rm HD}$  and working with the difference in  pH from the ${\rm{p}}K_a$ via $p = {\rm{pH}}-{\rm{p}}K_a$, the normalised excitation lifetime reads as
\begin{equation}
 \tilde{\tau}=   \frac{\tau({\rm{pH}})}{\tau_{\rm HD}} = \frac{1+10^{p}}{1+\gamma 10^{p}} \,.
\end{equation}
The carbon dot is most sensitive at the point of inflexion, where the slope with respect to pH (or $p$) is greatest. We note from Fig.~\ref{fig:pHlifetime} that the pH dependence of the lifetime is well characterised on a log scale.
 Hence the relevant point of inflexion is $d^2 \log_{10} \tilde{\tau}/ dp^2=0$ (rather than $d^2 \tilde{\tau}/ dp^2$), and is found at $p=-(1/2)\log_{10} \gamma$, that is, at $\pH_0 = \pK_a - 0.5\log_{10}( \Gamma_{\rm D}/\Gamma_{\rm HD})$. 
 We note that on a log scale, the slope $s=d\log_{10}\tilde{\tau}/dp$ is approximately constant over the region of pH sensitivity, such that $\log_{10}\tilde{\tau}$ is linear over the pH-sensitive region:
 \begin{equation}
 \label{eq:tau_linear}
     \log_{10}\tilde{\tau} \approx -\log_{10}\gamma  \left(\frac{\sqrt{\gamma}}{1+\sqrt{\gamma}}\right)
     + \left(\frac{1-\sqrt{\gamma}}{1+\sqrt{\gamma}}\right) \left(\pH -\pK_a\right)\,.
 \end{equation}
 The width of the pH-sensitive region can then be identified from the point where the line reaches $\log_{10}\tilde{\tau} = 0$ (the low pH plateau of the protonated state where $\tau=\tau_{\rm HD}$ and $\tilde{\tau}=1$) 
 and $\log_{10}\tilde{\tau} = \log_{10}({\tau_{\rm D}}/{\tau_{\rm HD}})$ (the high pH plateau of the deprotonated state where $\tau=\tau_{\rm D}$).
  Assuming $\gamma$ is large (or, equivalently, small), this  gives a width of the pH-sensitive region of $\log_{10}\gamma$ for a pH-sensitive interval
  \begin{equation}
      \pH \in [ \pK_a-\log_{10}\gamma ,\;\;   \pK_a ]\,.\label{eq:phrange}
  \end{equation}
Recalling $\gamma=\Gamma_{\rm D}/\Gamma_{\rm HD}$, we see that the sensor's sensitivity is maximised for small $\gamma$ (or, equivalently, large $\gamma$). Hence, the utility of the sensor is optimised when a large difference between the deprotonated and protonated rates is available. The conditions $\Gamma_{\rm D} \ll \Gamma_{\rm HD}$ (or, equivalently, $\Gamma_{\rm D} \gg \Gamma_{\rm HD}$), are desirable.

As a corollary, the pH-dependent change in fluorescence lifetime could be employed for titration of dye molecules for the purpose of measuring the acid constant $\pK_a$, though only in the case where $\gamma$ is small (or large); otherwise, the FL will remain independent of pH. The $\pK_a$ will correspond to the pH point where the fluorescence lifetime reaches the lower-lifetime plateau,  as indicated by arrows in Fig.~\ref{fig:pHlifetime}, \emph{not} the midpoint of the transition region.  
Values of $\pK_a$,  and $\tau_{\rm D}$ (via $\gamma$) can be obtained by fitting measured $\tau$ against the straight line in Eq.~\eqref{eq:tau_linear}, if $\tau_{\rm HD}$ has been identified from the low pH plateau. Alternatively $\tau_{\rm HD}$ may be obtained by fitting if $\tau_{\rm D}$ has been measured from the high pH plateau.

\section{Example: pH sensitivity of organic functional groups attached carbon dots}
We analyse the pH sensitivity of three dye molecules (m-phenylenediamine, phloroglucinol, disperse blue 1) motivated by the experiment reported in Ref.~\cite{szapoczka2023fluorescence}. The corresponding equilibrium constants $\mathrm{p}K_a$ for first protonation (or first deprotonation in the case of phloroglucinol) are summarized in table~\ref{tab:pkas}. The acid constant used for disperse blue 1 is that of the tethered molecule \cite{https://doi.org/10.1002/elan.200302902}, which is likely higher than that of the free molecule (related solway blue dye molecules \cite{PetersSumner1956} have $\pK_a < 1$). For comparison, experimental excitation lifetimes for deprotonated and protonated m-PD on the functionalised carbon dot  (CD4, which are based on m-PD, chosen due to linear behaviour) synthesised in Ref.~\cite{szapoczka2023fluorescence} using the linearised excitation lifetime~\eqref{eq:lin} are $\Gamma_{\rm D} = 0.353 \textrm{ ns}^{-1}$ and $\Gamma_{\rm HD} = 0.328 \textrm{ ns}^{-1}$, respectively. Note that these experimental rates are the response of the entire carbon dot. The separation approach introduced in Sec.~\ref{sec:internal}, separating the wavefunction of the functional group from the wavefunction of the carbon dot, cannot be applied to estimate these rates by only considering the functional groups since these carbon dots have a size distribution below 10~nm (wavelength of the confined electron~\cite{pelton2013introduction}). However, the rate mixing model itself~\eqref{eq:ratemixing}, respectively linearised~\eqref{eq:lin}, is more general and unaffected by this separation approach.

\begin{table}[htb]
    \centering
    \begin{tabular}{l|l}
        Molecule & \, ${\rm{p}}K_a$ \\\hline
        m-phenylenediamine \cite{pkas,10.1246/bcsj.60.4409}& \, 4.96 \\
        phloroglucinol \cite{doi:10.1021/ja00056a016} &  \, 8.9 \\
        disperse blue 1 (tethered)\cite{https://doi.org/10.1002/elan.200302902} & \, 5
    \end{tabular}
    \caption{Equilibrium constant $\mathrm{p}K_a$ (first acid constants) for the investigated molecules.}
    \label{tab:pkas}
\end{table}

The environmental impact of the host material (here: carbon dot) and the solvent has to be taken into account via the scattering Green function~\eqref{eq:Green} which can vary from continuous media\cite{10.1063/5.0106503} to atomic\cite{D2CP03349F} or molecular approaches,\cite{D0CP02863K} which can be used to describe more complex scenarios\cite{doi:10.1021/jacs.1c04880} or doping of the carbon dot.\cite{C5RA01986A} The atomic structure of the carbon dot influences the optical response of the carbon medium, which in turn provides an environmental shift in the pH-dependent excitation lifetime of a bound dye molecule.~\cite{YAO2019235,doi:10.1021/acssensors.9b00514} To illustrate the principle of the method,  we  consider a simple planar interface using the optical response of graphite to represent the carbon medium.

By treating the impact of the carbon dot itself via the environmental effects, we are assuming that the electronic wave functions of the carbon dot and the dye molecule can be separated. Hence, the theory is not applicable to quantum-dot-based \pH\ sensors,\cite{C5RA01986A} where the dot takes the role of the entire molecule.

We treat the residual dye centre of the functional group as a polarisable point particle bound at a certain distance $z$ from the surface of the carbon dot. The influence of the environment, both the carbon dot and aqueous medium, is expressed via reflection coefficients describing the change in dielectric function across the interface between two media. 

\begin{figure}[htb]
    \centering
    \includegraphics[width=5in]{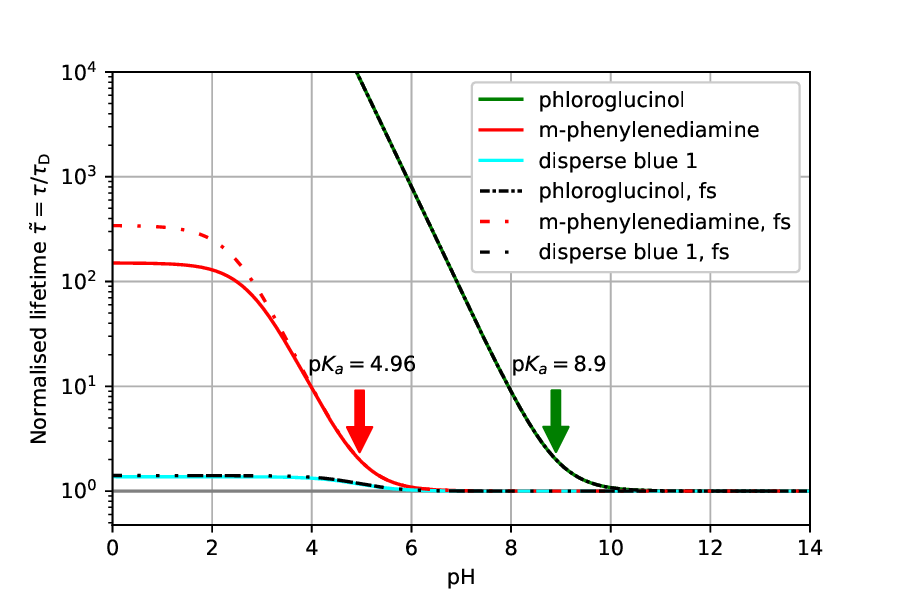}
    \caption{Change of excitation lifetimes at different \pH\ rescaled to the deprotonated lifetime $\tau_{\rm D}$ for m-phenylenediamine (red line), phloroglucinol (green line) and disperse blue 1 (blue line). The dashed-dotted lines illustrate the corresponding sensors without the impact of the carbon dots.}  
    \label{fig:pHlifetime}
\end{figure}

For large radii of the carbon dot, the electromagnetic scattering in such a system can be approximated by Fresnel reflection at a planar interface. If we simply embedded the functional group in the aqueous environment at distance $z$ from the carbon dot, the reflection coefficient for the relevant water-carbon interface would be negative since the dielectric function of carbon (both graphite \cite{doi:10.1142/8479} and diamond \cite{BERGSTROM1997125}) is higher than that of water \cite{doi:10.1021/acs.jpcb.0c00410} at optical/UV frequencies. This would result in an unphysical (or at least rare) increase in excitation lifetime due to the positiveness of the optical mode density~\cite{PhysRevLett.89.033001,PhysRevA.95.022509}.  
We, therefore, invoke a two-interface model placing the dye molecule in a (planar) vacuum cavity lying between the carbon and aqueous media. Reflection at a planar water-vacuum-carbon interface leads to the scattering Green function~\cite{Buhmann12a}
\begin{align}
\mathbf{G}(\mathbf{r},\mathbf{r}',\omega) = \frac{i}{8\pi^2} \int \frac{d^2 k^\parallel}{k_1^{\perp}} \exp\left(i \mathbf{k}^\parallel \cdot (\mathbf{r}-\mathbf{r}') + i k_1^\perp (z+z')\right) \left[ r_s \mathbf{e}_{s+}^1 \mathbf{e}_{s-}^1 + r_p \mathbf{e}_{p+}^1 \mathbf{e}_{p-}^1 \right], \label{eq:GreenPlanar}
\end{align}
with the Fresnel reflection coefficients
\begin{eqnarray}
r_s = \frac{k_1^\perp -k_2^\perp}{k_1^\perp+k_2^\perp}\,,\quad
r_p = \frac{\varepsilon_2 k_1^\perp -\varepsilon_1 k_2^\perp}{\varepsilon_2 k_1^\perp+\varepsilon_1 k_2^\perp}\,,
\end{eqnarray}
the wave vector parallel to the plane $\kk^\parallel \perp \ee_z$ and its component towards $z$ direction
$
k^\perp_j = \sqrt{\varepsilon_j  \omega^2/c^2-{k^\parallel}^2}
$. As $z$ describes the binding distance of the functional group to the carbon dot, the Green tensor~(\ref{eq:GreenPlanar}) simplifies in the nonretarded limit~\cite{10.1063/5.0037629}
\begin{eqnarray}
\IM\GGp(\rr,\rr,\omega)=\frac{c^2}{32\pi\omega^2 z^3}\IM \left[R(\omega)\right]  \begin{pmatrix}1 & 0 & 0\\
0 & 1  & 0 \\
0 & 0 & 2 
\end{pmatrix}\,.\label{eq:GP}
\end{eqnarray}
Here $R(\omega)$ is a multi-scattering reflection coefficient~\cite{doi:10.1021/acsearthspacechem.9b00019,PhysRevA.99.062512,C9CP03165K}
\begin{equation}
    R(\omega) = \frac{r_{\rm W}(\omega)r_{\rm C}(\omega) \mathrm e^{-{\rm i} \omega l /c}}{1-r_{\rm W}(\omega)r_{\rm C}(\omega) \mathrm e^{-{\rm i} \omega l /c }}\,,
\end{equation}
describing the influence of the two neighbouring interfaces on the dye molecule,
where $l$ is the thickness of the immediate surrounding vacuum layer. The non-retarded reflection coefficient for the vacuum-water and vacuum-carbon interfaces is
\begin{equation}
    r_i(\omega) = \frac{\varepsilon_{i}(\omega)-1}{\varepsilon_{i}(\omega)+1}\,,
\end{equation}
using dielectric functions for water and carbon, $\varepsilon_{\rm W}$ and $\varepsilon_{\rm C}$, respectively. Additionally, we consider local-field corrections addressing the propagation of photons through the liquid surrounding the carbon dot, applying Onsager's real cavity model,\cite{doi:10.1021/acs.jpca.7b10159,C9CP03165K,D3NR05396B} to modify the reflection coefficient\cite{doi:10.1021/acs.jpca.7b10159}
\begin{equation}
    R^\star(\omega) = R(\omega) \left(\frac{3\varepsilon_{\rm W}(\omega)}{1+2\varepsilon_{\rm W}(\omega)}\right)^2\,.
\end{equation}
By combining the Einstein coefficient~\eqref{eq:Einstein} and the medium-assisted change of transition rate~\eqref{eq:Gamma} together with the planar Green function~\eqref{eq:GP}, the transition rates for the protonated and deprotonated molecules near the carbon dot can be estimated
by summing the free-space rate (Einstein coefficient) and medium-assisted rate
\begin{equation}
    \Gamma_{\rm HD/D} = \Gamma_{\rm fs}^{\rm HD/D}+\Gamma_{\rm ma}^{\rm HD/D} \,,\label{eq:gam}
\end{equation}
with

\begin{align}
\Gamma_{\mathrm{fs}}^{\mathrm{HD/D}} &= \frac{\omega_{\mathrm{HD/D}}^3 |\bm{d}_{\mathrm{HD/D}}|^2}{3\hbar\pi\varepsilon_0c^3}, \label{eq:free} \\
\Gamma_{\mathrm{ma}}^{\mathrm{HD/D}} &= \frac{1}{2\pi\hbar\varepsilon_0 l_{\mathrm{HD/D}}^3} \mathrm{Im} \left[ \left(\frac{3\varepsilon_{\mathrm{W}}(\omega_{\mathrm{HD/D}})}{1+2\varepsilon_{\mathrm{W}}(\omega_{\mathrm{HD/D}})}\right)^2 \frac{r_{\mathrm{W}}(\omega_{\mathrm{HD/D}}) r_{\mathrm{C}}(\omega_{\mathrm{HD/D}}) e^{-i \omega_{\mathrm{HD/D}} l_{\mathrm{HD/D}} /c}}{1 - r_{\mathrm{W}}(\omega_{\mathrm{HD/D}}) r_{\mathrm{C}}(\omega_{\mathrm{HD/D}}) e^{-i \omega_{\mathrm{HD/D}} l_{\mathrm{HD/D}} /c}} \right] \nonumber \\
&\quad \times \left([d_{\mathrm{HD/D}}^x]^2 + [d_{\mathrm{HD/D}}^y]^2 + 2[d_{\mathrm{HD/D}}^z]^2\right), \label{eq:ma}
\end{align}

where we located the centre of the transition in the centre of the dye molecule $z=l_{\rm HD/D}/2$.
We use the dielectric function of water from Ref.~\cite{doi:10.1021/acs.jpcb.0c00410} and graphite from Ref.~\cite{doi:10.1142/8479} to evaluate this equation.
Transition dipole moments and excitation energies were computed by NWChem \cite{nwchem} using Time-Dependent Density Functional Theory (TDDFT). Geometries were first optimised by conventional Kohn-Sham DFT. In both cases, a B3LYP  functional \cite{b3lyp} was used with a def2-TZVP  basis set \cite{def2-TZVP}. TDDFT was configured to calculate the first singlet excitations, of which the first ($S_1 \leftarrow S_0$) is applied in this work. 
The thickness of the vacuum layer can be approximated via the diameter of the vacuum cavity surrounding each functional molecule. We took cavity volumes from NWChem calculations by applying the Conductor-Like Screening Model (COSMO)\cite{COSMO} to the molecules in water. By assuming a spherical shape, the corresponding diameter can be obtained $l = \sqrt[3]{6V/\pi}$. The results are given in Table~\ref{tab:diameters}.
The resulting transition rates for the functional group in free-space~\eqref{eq:free} and attached to the dissolved carbon dot~\eqref{eq:ma} are given in table~\ref{tab:rates} and illustrated in Fig.~\ref{fig:pHlifetime}.

\begin{table}[h]
    \centering
    \begin{tabular}{|c|c|c|c|}
      \hline 
    & {Molecule} & {Cavity volume $V\,(\text{\AA}^3)$} & {Cavity diameter $l\,(\text{\AA})$} \\ \hline
    \multirow{3}{*}{\rotatebox[origin=c]{90}{depr.}} 
    & m-pd & 92.672 & 5.61 \\ 
    & phloroglucinol & 94.771 & 5.66 \\ 
    & disperse-blue1 & 175.831 & 6.95 \\ \hline
    \multirow{3}{*}{\rotatebox[origin=c]{90}{prot.}} 
    & m-pd & 90.611 & 5.57 \\ 
    & phloroglucinol & 91.894 & 5.60 \\ 
    & disperse-blue1 & 175.230 & 6.94 \\ \hline
    \end{tabular}
    \caption{Molecular occupation volume $V$ of in water dissolved protonated and deprotonated precursor molecules (m-phenylenediamine, phloroglucinol, and disperse-blue 1) and the corresponding spherical cavity diameters $l$. }
    \label{tab:diameters}
\end{table}
\begin{table}[h]
 \centering
 \begin{tabular}{|c|c|c|c|c|}
 \hline
 & Molecule & fs rates $\Gamma_{\rm fs}\,(\mathrm{s}^{-1})$ & ma rates $\Gamma_{\rm ma}\,(\mathrm{s}^{-1})$ & $\tau_{\rm ma}\,(\rm{ns})$ \\
 \hline
 \multirow{3}{*}{\rotatebox[origin=c]{90}{depr.}} 
 & m-pd & $1.032 \times 10^{7}$ & $9.238 \times 10^{11}$ & 0.001\\
 & phloroglucinol  &  $1.253 \times 10^{5}$ & $1.512 \times 10^{10}$ & 0.066\\
 & disperse-blue 1 &  $1.385 \times 10^{7}$ & $4.160 \times 10^{12}$ & 0.00002\\
 \hline
 \multirow{3}{*}{\rotatebox[origin=c]{90}{prot.}} 
 & m-pd & $3.003 \times 10^{4}$ &  $6.132 \times 10^{9}$ & 0.163\\
 & phloroglucinol  & $4.308 \times 10^{-2}$ & $5.291 \times 10^{3}$ & $2\times 10^5$\\
 & disperse-blue 1 &  $9.826 \times 10^{6}$ & $3.028 \times 10^{12}$ & 0.0003 \\
 \hline
 \end{tabular}
 \caption{Resulting transition rates for the protonated and deprotonated molecules in free-space (fs rates) and the medium-assisted (ma rates) corrections according to Eq.~\eqref{eq:free} and Eq.~\eqref{eq:ma} 
 and the medium-assisted excitation lifetimes $\tau_{\rm ma} =1/\Gamma_{\rm ma}$
 for m-phenylenediamine, phloroglucinol, and disperse-blue 1.}
    \label{tab:rates}
\end{table}

Figure~\ref{fig:pHlifetime} illustrates the change of excitation lifetimes of the functionalised carbon dots (functionalised with the dye molecules (m-phenylenediamine, phloroglucinol and disperse blue 1) as a function of the pH of the surrounding solvent, shown relative to the excitation lifetime of the deprotonated species. Phloroglucinol shows sensitivity over the broadest pH range since it has the greatest difference between $\Gamma_{\rm D}$ and $\Gamma_{\rm HD}$ (Table~\ref{tab:rates}). 
However, the rates for protonated phloroglucinol are  determined from an almost vanishing transition dipole moment indicating the tracking of a forbidden transition, resulting in the large magnitude seen in the excitation lifetime.
It can be observed that lifetimes grow with decreasing pH, which agrees with the experimental observation.\cite{szapoczka2023fluorescence} However, the impacts of the functional groups m-phenylenediamine and phloroglucinol are overestimated, whereas the disperse blue 1 shows only a small pH dependency, discrepant from experimental observations. We attribute this discrepancy to the choice of electronic transition selected from the TDDFT simulation. The considered relaxation $S_1\mapsto S_0$  is a fast-decaying electronic excitation that directly decays into the ground state. In contrast, experiments consider luminescence decay, which couples to lower vibrational states and, thus, decays via a cascade, enhancing the lifetime. It is important to note, however, that treatment of these more complex fluorescence paths still follows the scheme introduced in Sec.~\ref{sec:mixing}, given the transition rates $\Gamma_{\ce{D}}$ and $\Gamma_{\ce{HD}}$ appropriate to these relaxation paths. 
\section{Conclusion}
In the manuscript, we introduced a theoretical model describing the pH dependence of dissolved particles. 
This model~\eqref{eq:GammaPH} leading to~\eqref{eq:ratemixing}, is valid for isolated carbon dots at low volume fractions.  But the theory can otherwise be applied to interpolate the excitation lifetime at different \pH\ values using the lifetimes of fully protonated/deprotonated species (at low pH/high pH limits)  found by any means, whether experimentally or theoretically.
By considering the protonation and deprotonation state of functional groups, the introduced model explains the trends in experimentally observed behaviour and qualitatively describes the dependencies on the material properties. 
By accounting for mole fractions in mixed multi-component liquids, the introduced approach could be extended to arbitrary solvent media.
By applying established local-field correction models, we reduced the numerical costs for modelling the carbon dot to the consideration of the functionalising dye molecule. 
By employing an assumption of separability of the dye molecule from the surrounding media, both carbon and water, the quantum chemistry calculations can be reduced to solely considering the functional groups. This approach would not be suitable for analysing small quantum dots ($< 10$ nm: wavelength of the confined electron~\cite{pelton2013introduction}), where the carbon particle essentially is the molecule. To theoretically reproduce results obtained experimentally, computation of the full set of transitions between excited states would be required, particularly the phonon coupling enabling single-triplet transitions. The described effects depend on the chemical structure of the carbon dot, both carbon core and functional group. The description of the core carbon dot could be improved by considering the distribution of carbon phases, e.g. using the proportion of $sp^2$ to $sp^3$ bonds in the core to construct a mixed graphite/diamond model of the carbon optical spectrum, or using an atomic description of the core to calculate its dielectric function. An important question is whether the functional group's molecular structure can still be identified with the free-space precursor molecule. In any case, a more rich model of the pH dependence can be constructed by extending the model to include higher (de-)protonation states of the functional group.  

\section*{Acknowledgments}
J.F. gratefully acknowledges support from the European Union (H2020-MSCA-IF-2020, grant number: 101031712). The work was financially supported by the Norwegian Research Council, project number 309612 - SFI Smart Ocean.

\bibliographystyle{unsrt}  
\bibliography{references}

\end{document}